\documentclass{amsart}
\usepackage{amsmath, amsthm}
\usepackage{enumerate}
\usepackage{graphicx}
\usepackage{amsfonts}
\usepackage{amssymb}

\usepackage{natbib}

\newtheorem{theorem}{Theorem}[section]

\newtheorem{lemma}[theorem]{Lemma}
\newtheorem{proposition}[theorem]{Proposition}

\newtheorem{definition}[theorem]{Definition}

\newtheorem{remark}[theorem]{Remark}
\newtheorem{example}[theorem]{Example}

\numberwithin{equation}{section}

\def\P{{\mathsf{P}}}
\def\F{{\mathcal{F}}}

\def\R{ {\mathbb R} }

\def\Ex{ {E} }

\def\P{ \mathsf{P} }

\title{A Heat Kernel Approach to Interest Rate Models}
\author{Jir\^o Akahori} 
\address{(Jir\^o Akahori) Ritsumeikan University}
\email{akahori@se.ritsumei.ac.jp}
\author{Yuji Hishida}
\address{(Yuji Hishida) Mizuho Security Co. Ltd.}
\email{yuji.hishida@gmail.com}
\author{Josef Teichmann}
\address{(Josef Teichmann) Vienna University of Technology}
\email{jteichma@fam.tuwien.ac.at}
\author{Takahiro Tsuchiya}
\address{(Takahiro Tsuchiya) Vienna University of Technology}
\email{suci@probab.com}

\thanks{This research is partially supported by 
Open Research Center Project for Private Universities: 
matching fund subsidy from MEXT, 2004-2008. 
The authors gratefully acknowledge support by 
the Austrian Science Fonds (FWF) through 
the research grant Y328 (START prize project)}
\begin{document}

\begin{abstract}
We construct default-free interest rate models 
in the spirit of the well-known Markov funcional models: our focus
is analytic tractability of the models and generality of the approach. 
We work in the setting of state price densities and construct models 
by means of the so called
propagation property. The propagation property can be found implicitly
in all of the popular state price density approaches, 
in particular heat kernels share the propagation property 
(wherefrom we deduced the name of the approach). 
As a related matter, an interesting property of 
heat kernels is presented, too.
\end{abstract}

\bibliographystyle{aer}
\maketitle

\begin{center}
{\bf Key Wordes}: Interest rate models, Markov-functional, state price density, heat kernel.
\end{center}

\section{Introduction}
The {\em heat kernel approach} (HKA for short), 
which will be presented and discussed in this paper, 
is a systematic method to construct state price densities 
which are analytically tractable. To be precise on analytic tractability, 
we mean with this notion that bond prices can be calculated explicitly, 
and that caps, swaptions or other derivatives on bonds can be 
calculated up to one integration with respect to the law of the underlying 
factor Markov process. 
Therefore such models can be easily calibrated to market data 
and are useful for pricing and hedging purposes, 
but also for purposes of risk management.

The original motivation of introducing the HKA was 
in modelling of interest rates with jumps. 
In the HJM framework \citep*{HJM}, 
the drift condition becomes quite complicated 
(see H. Shirakawa's pioneering work \citep*{Shi}, see also 
\citep*{AkaTsu:07} and references therein) 
if taking jumps into account, while in the spot rate approach, 
one will find it hard to obtain explicit expressions of the bond prices
(the affine class is almost the only exception).
In \citep*{AkaTsu:07}, the state price density 
approach\footnote{The state price density, which is sometimes called 
the pricing kernel, or the state price deflater, is, roughly speaking, 
a Radon Nikodym derivative multiplied by a stochastic discount factor. 
A brief survey of the state price density and 
its relation to the interest rate modelling will be given 
in section \ref{SPDA}.} is applied and 
by means of transition probability densities of some 
L\'evy processes explicit expressions of 
the zero coupon bond prices are obtained. 
The HKA is actually a full extension of the method, 
and thanks to the generalization 
its applications are now not limited to jump term structure models 
as we will show in the sequel.

Before the presentation of the theorem, we will give a brief survey of the 
state price density approaches such as the potential approach by L.C.G. Rogers \citep*{Rog} or 
Hughston's approach \citep*{FleHug} \citep*{HugRaf}, etc, in Section \ref{SPDA} and Section \ref{lreview}. 
Our models are within the class of {\em Markov functional models} proposed by P.~Hunt, J.~Kennedy and A.~Pelsser
\citep*{HKP}, and therefore compatible with their practical implementations. One of the contributions 
of the HKA to the literature could be to give a systematic way to produce Markov functional models.

As a whole, this paper is meant to be an introduction to the HKA, together with some examples. Topics from practical viewpoint like fitting to the real market, 
tractable calibration, or econometric empirical studies, are not treated in this paper. Nonetheless we note that
HKA is an alternative approach to practical as well as theoretical problems in interest rate modeling. 

\section{State price density approaches}

\subsection{State price density approach to the interest rate modeling}\label{SPDA}
We will start from a brief survey of state price densities. 
By a {\em market}, we mean a family of price-dividend pairs $ (S^i, D^i ) $, $ i \in I $, 
which are adapted processes defined on a filtered probability space $ (\Omega, \F,  \P ,\{ \F_t \}) $. 
A strictly positive process $ {( \pi_t )}_{t \geq 0} $ is 
a {\em state price density} with respect to the market 
if for any $ i \in I $, it holds that \begin{equation}\label{CFF}
S^i_t = \pi_t^{-1} \Ex [ \int_{t+}^\infty \pi_s dD^i_s | {\mathcal F}_{t} ], 
\end{equation}
or for any $ T  >  t $, 
\begin{equation}\label{CFF2}
S^i_t = \pi_t^{-1} \Ex  [ \pi_T S^i_T 
+ \int_{t+}^{T-} \pi_s dD^i_s | {\F}_{t} ].
\end{equation}
In other words, $ \pi_T/\pi_t $ gives a (random) discount factor of a cash flow $1$ at time $ T $ multiplied by
a Radon Nikodym derivative.

If we denote by $ P^{T}_{t} $ the market value at time $ t $
of zero-coupon bond with maturity $ T $, then the formula (\ref{CFF2}) gives 
\begin{equation}\label{SPD}
P^{T}_{t}= \pi_t^{-1} \Ex [ \pi_T| {\F}_{t} ].
\end{equation}
From a perspective of modeling term structure of interest rates,
the formula (\ref{SPD}) says that, given a filtration, each strictly positive process $ \pi $ generates 
an {\em arbitrage-free} interest rate model. On the basis of this observation, we can construct arbitrage-free
interest rate models. Notice that we do not assume $ \pi_t $
to be a submartingale, i.e. in economic terms \emph{we do not assume positive short rates}.

\subsection{The Flesaker-Hughston model, the Potential Approach and an approach by Wiener chaos}\label{lreview}

The rational log-normal model by Flesaker and Hughston \citep*{FleHug}
was a first successful interest rate model derived from the state price density approach. 
They put 
\begin{equation*}
\pi_t = A(t) + B(t) e^{\sigma W_t - \frac{\sigma^2}{2} t },
\end{equation*}
where $ A $ and $ B $ are deterministic decreasing process, 
$ W $ is a one dimensional standard Brownian motion and $ \sigma $ is a constant. 
The model is an analytically tractable, leads to positive-short rates, 
and is fitting any initial yield curve. 
Furthermore closed-form solutions for both caps and swaptions are available. 
Several extensions of this approach are known.

In \citep*{Rog}, L.C.G.~Rogers introduced 
a more systematic method to construct positive-rate models
which he called the {\em potential approach}. 
Actually he introduced two approaches;
in the first one, 
\begin{equation*}
\pi_t = e^{-\alpha t } R_\alpha g(X_t) / R_\alpha g(X_0), 
\end{equation*}
where 
$ R_\alpha $ is the resolvent operator of a Markov process 
on a state space $ S $,  
$ g $ is a positive function on $ S $, and $ \alpha $ is a positive constant. 
In the second one, $ g $ is specified as a linear combination of the eigenfunctions 
of the generator of $ X $. Note that when $ X $ is a Brownian motion, 
$ e^{\theta x} $ for any $ \theta \in \R $ is an eigenfunction of its generator, and 
the model gives another perspective 
to the rational lognormal models above. 

In \citep*{HugRaf} L.~Hughston and A.~Rafailidis proposed yet another framework for interest rate modelling
based on the state price density approach, 
which they call a {\em chaotic approach to interest rate modelling}. 
The Wiener chaos expansion technique is then used 
to formulate a systematic analysis of the structure and 
classification of interest rate models. 
Actually, M.R.~Grasselli and T.R.~Hurd \citep*{Gra} 
revisit the Cox-Ingersoll-Ross model of interest rates 
in the view of the chaotic representation and 
they obtain a simple expression 
for the fundamental random variable $X_{\infty}$ 
as $ \pi_t = \Ex [X_{\infty} - 
(\Ex [X_{\infty}|\mathcal{F}_t])^2 | \mathcal{F}_t] $ 
holds.
In line with these,  a stable property of the second-order chaos is shown in \citep*{AH}. 

\section{The Heat Kernel Approach}\label{HKA1}
In this section we present the simple concepts 
of the heat kernel approach (HKA) and several classes 
of possibly interesting examples. 
The guiding philosophy of HKA is the following: 
if one can easily calculate expectations of the form $ E(f(X_t)) $ 
for some factor Markov process $X$, and if one knows \emph{explicitly} 
one additional function $ p $ depending on time and the state variables, 
then one can construct explicit formulas for bond prices, 
and analytically tractable formulas for caps, swaptions, etc 
(i.e., those formulas are evaluated by one numerical integration 
with respect to the law of the underlying Markov process). 
In other words, any explicit solution of a well-understood 
Kolmogorov equation with respect to some 
Markov process leads to a new class of interest rate models.
\subsection{Basic concept}
We consider a general Markov process $ {(X^x_t)}_{t \geq 0,x \in S} $ 
on a polish state space $ S $ and a probability space 
$ (\Omega,\mathcal{F},P) $ 
with filtration $ {(\mathcal{F}_t)}_{t \geq 0} $. 
\begin{definition}
Let $ p(t,x) $ defined on $ \mathbb{R}_{\geq 0} \times S $ 
be a strictly positive function. 
It is said to satisfy the {\bf propagation property with respect to $X$} if
\begin{equation}\label{propro}
E(p(t,X^x_s))=p(t+s,x)
\end{equation}
for all $ t > 0, s \geq 0 $ and $ x \in S $.  
We assume furthermore that 
$ p(t,X_t) \in L^1(\Omega) $ for $ t > 0 $.
\end{definition}
Let $p$ satisfy the propagation property 
with respect to the Markov process $X$ and
$ p (0,x) \in L^1 (\R) $. 
We define then a state price density $ \pi_t $ through
\begin{equation}\label{SPDofHKA}
\pi_t := p(t,X^x_t).
\end{equation}
We clearly have the following basic assertion:
\begin{theorem}\label{HKAbond}
Bond prices are given through
$$
P^T_t = \frac{p(2T - t,X^x_t)}{p(t,X^x_t)},
$$
for $ 0 \leq t \leq T $ and $ x \in S $.
\end{theorem}
\begin{proof}
The bond prices are calculated with respect to the state price density 
(\ref{SPDofHKA}) through
$$
P^T_t = \frac{1}{\pi_t}E(\pi_T | \mathcal{F}_t).
$$
Actually, by the Markov property and by the propagation property 
we can calculate the right hand side,
\begin{align*}
E(\pi_T | \mathcal{F}_t) & = E(p(T,X^x_T) | \mathcal{F}_t) \\
& = E(p(T,X^y_{T-t}))|_{y=X^x_t} \\
& = p(2T - t, X^x_t ) 
\end{align*}
for all $ x \in S $ and $ 0 \leq t \leq T $.
\end{proof}
\begin{remark}
Notice that this remarkable simple formula is given with respect 
to the \textbf{historical}/\textbf{physical} measure.
\end{remark}
A derivative with payoff $ G $ on $ P^T_t $ for some $ 0 \leq t \leq T $ has at time $ t = 0 $ the price
$$
\frac{1}{\pi_0}E \bigl[p(t,X^x_t) G\big(\frac{p(2 T - t,X^x_t)}{p(t,X^x_t)}\big) \bigr],
$$
which is analytically tractable if one knows how to calculate expectations of the form $ E(f(X^x_t) $.
\begin{remark}
Notice that the initial term structure is given in the previous setting through
$$
P(0,T) = \frac{p(2 T ,x)}{p(0,x)}
$$
for $ T \geq 0 $.
\end{remark}

\subsection{Generic example}\label{genericexample}
Let $ {(X^x_t)}_{t \geq 0,x \in S} $ with a polish state space $ S $ on a probability space $ (\Omega,\mathcal{F},P) $ with filtration $ {(\mathcal{F}_t)}_{t \geq 0} $.
Take a measurable, bounded $ h: S \to \mathbb{R}_{\geq 0} $, then
\begin{equation}\label{generic}
p(t,x):= E(h(X^x_t))
\end{equation}
satisfies the propagation property (\ref{propro}).

This example demonstrates that the previous theory covers 
the results of \citep*{AkaTsu:07}. 
Indeed, consider a $ d $-dimensional L\'evy process $ {(X_t)}_{t \geq 0} $ starting at $ 0 $ in its natural filtration. 
We assume that it has a density $y \mapsto p(t,y) $ 
with respect to Lebesgue measure. Then $ X^x_t := x - X_t $ defines a Markov process on 
$ \mathbb{R}^d $ and we have that
\begin{equation}\label{pp0}
E(p(t,x-X_s)) = p(t+s,x)
\end{equation}
for $ s,t>0 $ and $ x \in \mathbb{R}^d $. 

Notice that this example can be associated to the previous Example \ref{genericexample} 
if one allows $ h = \delta_0 $ in the sense of distributions.

\subsection{A generic affine example} Consider an affine Markov process $ X $ on 
$ \mathbb{R}^n_{\geq 0} \times \mathbb{R}^m $ and $ \mu_1,\ldots,\mu_r \in \mathbb{R}^{n+m} $ such that
$ E(\exp(\langle \mu_i ,  X_t \rangle)) $ exist. Then we define
$$
h(x)= \sum_{i=1}^r a_i \exp(\langle \mu_i ,  x \rangle))
$$
for some numbers $ a_i > 0 $. In the spirit of the previous generic example we obtain
$$
p(t,x)= \sum_{i=1}^r a_i E(\exp(\langle \mu_i ,  X_t \rangle))  
$$
for $ t \geq 0 $ and each initial value $ x \in \mathbb{R}^n_{\geq 0} \times \mathbb{R}^m $. 
Notice that the ``Laplace transforms'' $E(\exp(\langle \mu_i ,  X_t \rangle))$ are easily calculated for affine processes, which yields a concrete representation of the prices in the physical measure. These models are completely new interest models constructed from affine models. For instance, the derived short rate models have interesting (econometric) features. An example driven by the affine Cox-Ingersoll-Ross process is visualized in Figure 1.

\begin{figure}
\begin{center}
\includegraphics*[width=8cm,height=5cm]{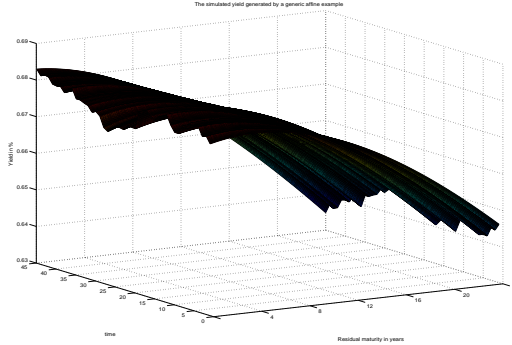}
\end{center}
\caption{Simulated Yields in the generic affine class driven by CIR}
\end{figure}

\subsection{Pricing formula of Swaptions}
The pay-off (at its maturity $ T_\alpha $) of a (unit of) swaption is
\begin{equation}\label{pay-off1}
(S_{\alpha, \beta} (T_\alpha) - K )^+ 
\sum_{i = \alpha +1}^\beta \tau_i P (T_\alpha, T_i ),
\end{equation} 
where $ \{ T_\alpha, T_{\alpha +1}, \ldots, T_{\beta} \} $ with $ \tau_i = T_i - T_{i-1} $ is 
the tenor structure of the swap contract, and 
\begin{equation*}
S_{\alpha, \beta} (T_\alpha)
= \frac{1 - P(T_\alpha, T_\beta)}
{\sum_{i = \alpha +1}^\beta \tau_i P (T_\alpha, T_i )}
\end{equation*}
is the swap rate (see e.g. \citep*{BriMer:06}*{Chapter 1, section 6}).
Note that (\ref{pay-off1}) can be rewritten as
\begin{equation}\label{pay-off2}
\{ 1 - P(T_\alpha, T_\beta) 
- K \sum_{i = \alpha +1}^\beta \tau_i P (T_\alpha, T_i ) \}^+. 
\end{equation}
Therefore in the state price density approach, the quantity
\begin{equation*}
\mathbf{sw}_t := \frac{1}{\pi_t} 
E (\pi_{T_\alpha} \{ 1 - P(T_\alpha, T_\beta) 
- K \sum_{i = \alpha +1}^\beta \tau_i P (T_\alpha, T_i ) \}^+ 
| \mathcal{F}_t )
\end{equation*}
gives a fair price of the swaption. Substituting 
$$ 
P ( T_\alpha, T_i ) 
= E ( \pi_{T_i} | \mathcal{F}_{T_\alpha}) / \pi_{T_\alpha},
$$
we get
\begin{equation*}
\mathbf{sw}_t =  \frac{1}{\pi_t} 
E ( \{ \pi_{T_\alpha}  - E (\pi_{T_\beta} | \mathcal{F}_{T_\alpha})
- K \sum_{i = \alpha +1}^\beta \tau_i  
E (\pi_{T_i} | \mathcal{F}_{T_\alpha}) \}^+ | \mathcal{F}_t ). 
\end{equation*}
By HKA, namely
\begin{equation*}
\pi_t = p (t, X_t), 
\end{equation*}
where $ p $ has the propagation property w.r.t. a Markov process $ X $.
We have
\begin{equation}\label{swaption:formula}
\begin{split}
\mathbf{sw}_t & =  \frac{1}{p (t, X_t)} \times \\
& \quad \times E^{X_t} [ \{ p ({T_\alpha}, X_{T_\alpha -t})
 - p ( 2T_\beta - T_\alpha, X_{T_\alpha -t}) - \\
& \qquad - K \sum_{i = \alpha +1}^\beta \tau_i  
p ( 2T_i - T_\alpha, X_{T_\alpha - t})\}^+],
\end{split}
\end{equation}
which is again -- up to integration with respect to the law of $ X $ -- explicit.
\subsection{Relations to short rate models}
Assume that $ p(t,x) $ is differentiable with respect to $t$. Then the short rate $ R_t = - \partial_T \log P^T_t |_{t=T} $ 
of the previous interest rate model is given through
$$
R_t = - \frac{1}{p(t,X^x_t)} 
\frac{\partial}{\partial t} 
p(t,X^x_t).
$$
Since $ p > 0 $ the sign of the short rate is determined through 
$$ 
- \frac{\partial}{\partial t} p(t,X^x_t).
$$
Notice that we do not necessarily have a short rate process.

\subsection{Eigenfunction models}\label{EigenModel}
Take a finite dimensional manifold $ M $ and take $ X $ a Markov process with values in $ M $. A non-vanishing function $ g :M \to \mathbb{R} $
with
\begin{equation}\label{EigenF}
E(g(X^x_t))= \exp(\mu t) g(x)
\end{equation}
for some $ \mu \leq 0 $ is called eigenfunction of $X$. Let $ g_i : M \to \mathbb{R} $, $ i =1,\cdots,N $ be eigenfuntions with eigenvalues $ \mu_i < 0 $, and let $ A_i :\mathbb{R}_{\geq 0} \to \mathbb{R}_{>0} $ be decreasing functions. Set
\begin{equation*}
v (t,x) := \sum_{i=1}^N A_i (t) g_i (x),
\end{equation*}
then we notice that $ v (t, X_t ) $ is a supermartingale, too. If $ v $ is strictly positive and $ N >1 $, then we can construct 
an arbitrage-free positive interest rate model by setting $ \pi_t := v (t, X_t ) $ as a state price density. 
We call such models the {\em eigenfunction models}.

\begin{remark}
The class includes the generic approach II
in the potential approach by Rogers \citep*{Rog},
and is included in the Flesaker-Hughston's general framework 
since $ e^{-\mu t} g (X_t) $ becomes a martingale (see section \ref{lreview}). 
\end{remark}

\subsection{Swaption Formula via eigenfunction models}

The formula (\ref{swaption:formula}) 
becomes extremely simple when we use the {\em eigenfunction model}. 
\begin{equation}\label{EigenS}
\pi_t = p (t, X^x_t) 
=1+ e^{\mu t } g (X^x_t) 
\end{equation}
where $ g $ satisfies (\ref{EigenF}). By applying (\ref{EigenS})
on (\ref{swaption:formula}), we obtain the following result. 
\begin{theorem}\label{swpformula}
The pricing formula of the swaption is 
\begin{equation*}
\mathbf{sw}_t =  \frac{1}{1+ e^{\mu t } g (X^x_t)} 
\int_M ( A g (y) - B )^+ \, 
P ( X_{T_\alpha -t}^x \in dy \,|\, x= X^x_t ),  
\end{equation*}
where 
\begin{equation*}
A= e^{\mu {T_\alpha}}
-e^{\mu (2 T_\beta - T_\alpha)}
- K \sum_{i = \alpha +1}^\beta \tau_i 
e^{\mu ({T_i}+ T_\beta - T_\alpha)},
\end{equation*}
and 
\begin{equation*}
B = K (T_\beta - T_\alpha)  .
\end{equation*}
\end{theorem}

\begin{example}[Brownian motion with drift] 
Let $X$ be a Brownian motion with drift: whose generator is ${\Delta}/2 - \kappa \cdot \nabla $ 
where $ \kappa \in \mathbb{R}^d $. For $ g(x) = e^{\langle c, x \rangle }$ with $ c \in \mathbb{R}^d $,
we have
\begin{equation*}
( {\Delta}/2 - \kappa \cdot \nabla ) g 
= ( |c|^2/2 - \langle \kappa , c  \rangle) g.
\end{equation*}
Therefore, if $ c $ is such that
$  |c|^2/2 - \langle \kappa , c  \rangle < 0 $
(this is always possible, take $  c = \kappa $ for example),
then the state price density
$$ p (t, X_t) = 1 + e^{\nu t  } g (X_t ), $$
where $ \nu := |c|^2/2 - \langle \kappa , c  \rangle $, 
gives a positive rate model. 
In this case \begin{equation*}
\begin{split}
& \mathbf{sw}_t = 
\frac{1}{1+ e^{\mu t } e^{\langle c, X_t^x \rangle}}
{\bf BS} \Bigl( 
Ae^{\langle c, X^x_t \rangle}, B, |c|\sqrt{T_\alpha -t}, 
\langle c,\kappa \rangle \sqrt{T_\alpha -t}\Bigr),
\end{split}
\end{equation*}
where 
$$ {\bf BS} (a, b, c, d) =
\frac{1}{\sqrt{2\pi}}
\int_\R ( a e^{cx+d} - b)^+ 
e^{-\frac{x^2}{2}}\,dx.
$$
Notice that the Black-Scholes formula for the price at time $ t $ of the call option 
with the strike $ K $ and the maturity $ T $ is given by $$ e^{-r (T-t)}
{\bf BS} (S_t, K, \sigma \sqrt{T-t}, (r-\frac{\sigma^2}{2})(T-t)). $$
\end{example}

\begin{remark}
This example corresponds to 
the Flesaker-Hughston's rational log-normal model \citep*{FleHug}.
\end{remark}

\begin{example}[squared OU process]
Let $X$ be an Ornstein-Uhlenbeck process on $ \R^d $, whose generator $ \mathcal{A} $ is given by 
\begin{equation*}
\mathcal{A}:= -\mu x \nabla  + \frac{1}{2}  {\Delta}
\end{equation*}
where $ \mu <0  $.
Setting $g(x)=e^{ \mu |x|^2}$, we have
\begin{equation*}
\mathcal{A}\,g(x) = \mu d \,g(x)
\end{equation*}
Therefore we have that $g$ is an eigenfunction with the eigenvalue ${\mu}d$.
Setting $ d=1 $ and $ p(t,x) = 1+ e^{\mu t } g (x) $, we obtain 
\begin{equation}\label{E1}
\begin{split}
& \mathbf{sw}_t = \frac{
A e^{\mu (T_\alpha -t) } g (X^x_t) 
\{ \Phi (d_1^+) - \Phi (d^-_1) \}
- B \{ \Phi (d_2^+) - \Phi (d_2^-) \} }
{1+ e^{\mu t} g (X^x_t) },
\end{split}
\end{equation}
where $ \Phi (x) = \int_{-\infty}^x 
\frac{ e^{-\frac{x^2}{2}}}{\sqrt{2 \pi}}\,dx $, 
\begin{equation}\label{E2}
d_1^{\pm} =\frac{\pm \sqrt{\frac{1}{\mu}
\log \frac{B}{A}} - 
X^x_t e^{\mu(T_\alpha -t)} }{\sqrt{
(2\mu)^{-1} (e^{2\mu (T_\alpha -t)} -1)}}
\end{equation}
and
\begin{equation}\label{E3}
d_2^{\pm} = \frac{\pm \sqrt{\frac{1}{\mu}
\log \frac{B}{A}} - X^x_t e^{-\mu(T_\alpha -t)} }{\sqrt{
(2\mu)^{-1} (1 -e^{-2\mu (T_\alpha -t)})}}.
\end{equation}
For the proof see the Appendix \ref{Ap:OU}.
\end{example}

\section{HKA and Positive Rate Models}\label{Theta}
As we have pointed out the propagation property of $ p $ does not ensure that 
the process $ p (t, X_t) $ is a supermartingale. Thus, the interest rate model with $ \pi_t = p (t,X_t) $
does not exclude the possibility of negative rates. In this section we provide models based on HKA and leading to positive short rates.

\subsection{Weighted heat kernel approach}
Let $ p $ be a positive function with propagation property with respect to a Markov process $ X $,
and a weight $ f : \R_{\geq 0} \times \R_{ \geq 0} \to \R_{>0} $ such that $ f (t,u-s) \leq f (t-s, u) $ 
for arbitrary $ t, u \in \R_{\geq 0} $ and $ s \leq t \wedge u $. Set
\begin{equation*}
q (t, x) = \int_0^\infty p(s,x) f(t,s) \,ds,
\end{equation*}
Then we have
\begin{proposition}\label{potential1}
The process $ q (t, X_t) $ is a supermartingale. 
\end{proposition}
\begin{proof}
For $ t > s $ and $ x \in S $, we have
\begin{equation*}
\begin{split}
E (q(t, X^x_t)|\mathcal{F}_s ) 
&=
\int_0^\infty E  ( p(u, X^x_t) | \mathcal{F}_s) f(t,u) \,du \\
&= \int_0^\infty p(u+ t-s, X^x_s)  f(t,u) \,du \\
& = 
\int_{t-s}^\infty p(v, X^x_s)  f(t, v-t+s) \,dv
\end{split}
\end{equation*}
Here we have used the propagation property of $ p $. 
Since
\begin{equation*}
\begin{split}
\int_{t-s}^\infty  p(v, X^x_s)  f(t, v-t+s) \,dv 
& \leq \int_{t-s}^\infty p(v, X^x_s)  f(t-(t-s), v) \,dv \\
& \leq  \int_0^\infty p(v, X^x_s)  f(s, v) \,dv 
= q (s, X_s^x), 
\end{split}
\end{equation*}
we have the desired result. 
\end{proof}
Thus interest rate models with $ \pi_t = q(t,X_t) $ have positive short rates. The class is an extension of the potential models defined in \citep*{Rog:04},
since we can take $ p $ as in \eqref{generic} and $ f (t,s) = e^{-\alpha (t+s)} $. 

\subsection{Killed Heat kernel approach}
Let $ V $ be a non-negative, measurable function defined on $ S $. Put 
\begin{equation*}
 q (t, x ) =  E ( \exp{ ( - \int_0^t V (X^x_s) \,ds ) } ), 
\end{equation*}
and
\begin{equation}\label{DB}
P_t^T = \frac{q ( 2T-t,  X^x_t )}{q(t,X^x_t)}.
\end{equation}
Then the bond market spanned by $ P^T $ is arbitrage-free with respect to the state price density
$$
\pi_t = q(t,X^x_t) \exp(-\int_0^t V (X^x_s) \,ds ).
$$
Indeed, this corresponds to an additional factor $ Y_t^y = y + \int_0^t V (X^x_s)\,ds $ for $ y \geq 0 $ added to the Markov process $X$. Since $ (X^x, Y^y ) $ is again a Markov process on $ S \times \mathbb{R}_{\geq 0} $, we notice that
\begin{equation*}
p(t, x, y) := E (e^{-Y^y_t}) = e^{-y} E ( \exp{( - \int_0^t V (X^x_s)\,ds )} ) = e^{-y} q(t,x) 
\end{equation*} 
defined on $ \mathbb{R}_{\geq 0} \times S \times \mathbb{R}_{\geq 0} $ satisfies the propagation property. Therefore the previously described bond prices are arbitrage-free with respect to the stated state price densities for initial values $ (x,0) $. Since $ q $ is decreasing in $t$ the short rates are again positive.

\subsection{Trace approach}\label{Trace}
We give a third, more involved method to obtain positive rate models. Though it is only valid for symmetric L\'evy processes at this stage, again the class is completely new. Let ${(X_t)}_{t \geq 0} $ be a $d$-dimensional L\'evy process, which is symmetric in the sense that $ X_1 \overset{\mathrm{d}}{=} - X_1 $. We define $ X^x_t = x+ X_t $. Note that in this case its L\'evy symbol is positive and symmetric, i.e.
\begin{equation*}
{E} (e^{2 \pi i \langle \xi, X_t^0 \rangle} ) = e^{- t\psi (\xi) }
\end{equation*}
for some 
\begin{equation*}
\psi : \mathbb{R}^d \to \mathbb{R}_{\geq 0},
\end{equation*}
which is symmetric in the sense that 
$ \psi (\xi) = \psi (-\xi) $.

Let $ \mu $ be a measure on $ \mathbb{R}^d $ 
such that
\begin{equation*}
\mu (-A) = \mu (A) 
\end{equation*}
for any Borel set $ A $. 
Put 
\begin{equation*}
u (t,x) =\int_{\mathbb{R}^d} e^{2\pi i \langle \eta, x \rangle} 
e^{- t \psi (\eta)}
\mu (d\eta),
\end{equation*}
assuming it is finite (except at $ t=0 $) for each $ x \in \mathbb{R}^d $. 
\begin{lemma}
The function $ u $ is non-negative and has the propagation property with respect to 
$ \{ X^x \} $. 
\end{lemma}

\begin{proof}
First, notice that by the symmetry of $ \mu $ we have
\begin{equation}\label{realU}
u (t,x) =\int_{\mathbb{R}^d} \cos( 2 \pi \langle \eta, x \rangle)  
e^{- t \psi (\eta)}
\mu (d\eta) \in \mathbb{R},
\end{equation}
and $ u (t,x) = u (t,-x) $ for any $ x \in \mathbb{R}^d $. Then by Bochner's theorem we know that 
$ u (t, \cdot) $ is non-negative definite and we see that 
$ u (t, x) \geq 0 $ for any $ x \in \mathbb{R}^d $.

Since $ | u_t (x) | \leq 1 $, $ \mathbf{E} [u(s, X^x_t) ] \leq 1 $ for any
$ s, t > 0 $ and $ x \in \mathbb{R}^d $, we have \begin{equation*}
\begin{split}
E ( u(s, X^x_t) )
&= E \left( \int_{\mathbb{R}^d} 
e^{2\pi i \langle \eta, X_t^x \rangle} e^{- s \psi (\eta)}
\mu (d\eta)
\right) \\
&= \int_{\mathbb{R}^d} 
{E}  ( e^{2\pi i \langle \eta, X_t^x \rangle} ) e^{- s \psi (\eta)}
\mu (d\eta) \\
&= \int_{\mathbb{R}^d} e^{2\pi i \langle \eta, x \rangle}
E  ( e^{2\pi i \langle \eta, X_t^0 \rangle} ) e^{- s \psi (\eta)}
\mu (d\eta) \\
&= \int_{\mathbb{R}^d} e^{2\pi i \langle \eta, x \rangle}
e^{- (s+t) \psi (\eta)}
\mu (d\eta) = u (t+s, x). 
\end{split}
\end{equation*}
\end{proof}

\begin{remark}
If the Fourier transform 
\begin{equation*}
f (x) := \int_{\mathbb{R}^d} e^{ 2 \pi i \langle x, \xi \rangle} 
\mu (d \xi) 
\end{equation*}
exists and is bounded in $ x $, then by the symmetry of $ \mu $ it holds that $ f (x) \geq 0 $, and $ u $ has the following expression:
\begin{equation*}\label{positivedefinite}
u (t,x) = E ( f (X^x_t)) .
\end{equation*}
On the other hand, let $ \mu $ be the Lebesgue measure of $ \mathbb{R}^d $. If the density $ p_t (x) $ 
of $ X^0_t $ exists in $ L^1 \cap L^2 $, we have
\begin{equation*}
u (t,x) = p_t (x). 
\end{equation*}
That is, $ u $ is the heat kernel. 
\end{remark}

We are now able to prove the following theorem on supermartingales constructed from heat kernels:
\begin{theorem}\label{mainTh}
Let $ c $ be a positive constant greater than $ 2 $ and let $ \lambda $ be any positive constant. Set
$\pi_t := u( \lambda + t,X^x_{t}) + c u ( \lambda + t,0)$. Then $\pi$ is a supermartingale with respect to 
the natural filtration of $ X $. 
\end{theorem}

\begin{proof}
By the expression (\ref{realU}), we have 
\begin{equation}\label{eq1}
\begin{split}
& u( \lambda + t,X^x_{t}) - u( \lambda +2T-t,X^x_{t}) \\
& \hspace{2cm} = 
\int_{\mathbb{R}^d}
\cos 2 \pi \langle X^x_{t} , \xi \rangle 
(e^{- (t+\lambda) \psi(\xi)} 
- e^{- (\lambda +2T-t) \psi(\xi)} ) \mu(d \xi) 
\end{split}
\end{equation}
Since $\psi \geq 0$ and $\lambda$ is positive, 
\begin{align*} 
& \text{RHS of (\ref{eq1})} \\
&\geq \int_{\mathbb{R}^d}
	(e^{- (t+\lambda) \psi(\xi)} 
	- e^{- (\lambda+2T-t)) \psi(\xi)}) \mu(d \xi) \\
	&= -(u(t+\lambda,0) - u(\lambda + 2T-t, 0)).
\end{align*}
Thus we obtain  
\begin{align*} 
& u(t+\lambda,X^x_{t}) +u (t+\lambda,0) \\
& \hspace{2cm} \geq 
u(\lambda + 2T-t,X^x_{t})+ u( \lambda + 2T-t, 0),
\end{align*}
and then  
\begin{equation}\label{2ndineq}
\begin{split} 
& u (t+\lambda,X^x_{t}) +c u (t+\lambda,0) \\
&\geq u ( \lambda + 2T-t, X^x_{t} )+ u (\lambda + 2T-t, 0)
+ (c-1) u (t+\lambda,0). 
\end{split}
\end{equation}
On the other hand, by an elementary inequality we have 
\begin{align*} 
& u (\lambda + 2T-t, 0) +  u (t+\lambda,0) \\
&= 
\int_{\mathbb{R}^d}
(e^{- (\lambda + 2T-t) \psi(\xi)} 
+ e^{- (t+\lambda) \psi(\xi)}) \mu(d \xi) \\
&\geq 2
\int_{\mathbb{R}^d}
e^{- \frac{1}{2} (\lambda + \lambda + 2T) \psi(\xi)} \mu(d \xi), 
\end{align*}
and since $\lambda$ is positive, we obtain 
\begin{equation}\label{3rdineq}
\begin{split}
& u (\lambda + 2T-t, 0) +  u (t+\lambda,0) \\
& \hspace{2cm} \geq 2 
\int_{\mathbb{R}^d}
e^{-  (T+ \lambda) \psi(\xi)}\mu(d \xi)
= 2 u ( T +\lambda, 0 ). 
\end{split}
\end{equation}
Combining (\ref{2ndineq}) and (\ref{3rdineq}), 
we have
\begin{equation}\label{4thineq}
\begin{split}
& u (t+\lambda,X^x_{t}) +c u (t+\lambda,0)- \\
& \qquad - u (\lambda + 2T -t, X^x_t ) - c u (T + \lambda , 0 ) \\
& \hspace{2cm} 
\geq (c-2) ( u (t+ \lambda, 0) - (c-2) u (T + \lambda, 0 ) ).
\end{split}
\end{equation}
The right-hand-side of (\ref{4thineq}) is non-negative 
since $ c \geq 2 $ and 
\begin{equation*}
u (t,0) = \int_{\mathbb{R}^d} e^{- t \psi (\xi) } \, \mu (d \xi)
\end{equation*}
is decreasing in $ t $. 
Finally, since we have 
$$
E[\pi_T | \mathcal{F}_t] = 
u (\lambda+2T-t, X^x_{t}) +  c u (T+\lambda,0), 
$$
we now see that $\pi_t$ is a supermartingale. 
\end{proof}

\subsection{Examples}\label{TM}
By the above theorem, we can now construct a positive interest rate model 
by setting $ u (\lambda +2T-t, X^x_{t}) +  c u (T+\lambda,0) $
to be a state price density, or equivalently 
\begin{equation*}
P^T_t = \frac{u(\lambda + 2T-t,X^x_{t}) + c u(T+\lambda,0)}
	{u(t+\lambda,X^x_{t}) + c u(t+\lambda,0)}.
\end{equation*}
Here we give some explicit examples. In all examples, $ c $ is a constant greater than $ 2 $ and
$ \lambda $ some positive constant. 

\begin{example}[Heat Kernel with Trace]
A simple example is obtained by letting 
\begin{equation*}
u(t,x) = \frac{1}{\sqrt{2 \pi t}} e^{- \frac{x^2}{2t}},
\end{equation*}
where the bond prices are expressed as 
\begin{equation*}
P^T_t = 
\frac{
\sqrt{\frac{\lambda + t}{\lambda + 2 T-t}} e^{- \frac{W^2_t}{\lambda_T + 2T -t} } 
	+ 
\sqrt{\frac{\lambda + t}{\lambda + T}}c
}
{
e^{- \frac{W^2_t}{\lambda + t} } + c
}
\end{equation*}
\end{example}

\begin{example}[Quadratic Gaussian with Trace]
A similar model can be obtained by 
setting, for $ \alpha > 0 $, 
\begin{equation*}
\begin{split}
u (t,x) &= \frac{1}{\sqrt{( t \alpha^2+1)}}
e^{-\frac{\alpha^2}{2(t \alpha^2 +1)} x^2} \\
&\left( =E_{x} (e^{- \frac{\alpha^2}{2} W^2_t}) = 
E \left( \int_\mathbb{R} e^{i y \alpha W_t^x} 
\frac{1}{\sqrt{2 \pi}}e^{ - \frac{y^2}{2} } \,dy \right) 
\right).
\end{split}
\end{equation*}
A visualization of this model is shown in Figure 2.
\begin{figure}
\begin{center}
\includegraphics*[width=8cm,height=5cm]{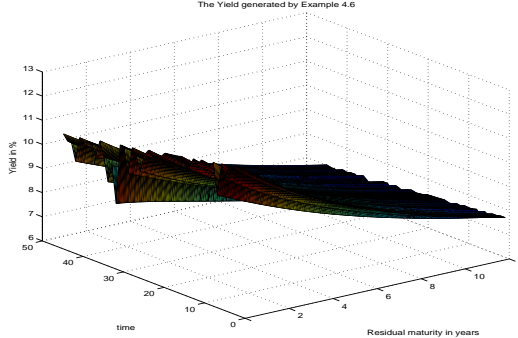}
\end{center}
\caption{Simulated Yields for the trace class model driven by a Quadratic Gaussian}
\end{figure}

\end{example}

\begin{example}[Cauchy $0$ with Trace]
Let $Z^x$ be a Cauchy process in $ \R^d $ starting from $ x $. The explicit form of the transition density of $ Z^0 $ is known to be
\begin{equation}\label{CauchyD}
p(t,x) :=
\frac{ \Gamma ((d+1)/2) \theta t }
{  ( \pi \{ (\theta t )^2+|x- t \gamma  |^2 \})^{(d+1)/2}},
\end{equation}
where $ \theta > 0 $ and $ \gamma \in \R^d $ (see e.g.\citep*{Sat}).
Note that Cauchy processes are strictly stable (or self-similar) process with parameter $ 1 $;
namely, 
\begin{equation*}
Z_{a t} \overset{\mathrm{d}}{=} a Z_t, \quad a > 0, t > 0.
\end{equation*} 
This property is seen from their L\'evy symbol: actually we have
\begin{equation*}
\Ex [e^{i \langle \xi, Z_t \rangle }] 
= e^{ - t ( \theta |\xi| - i \langle \xi, \gamma \rangle )}.
\end{equation*}
If we set $ \gamma = 0 $, the density turns symmetric and therefore, setting $ \pi_t = p (\lambda + t, Z^x_t ) + c p (\lambda + t, 0) $,
we obtain an explicit positive term structure model. We give the explicit form of the bond prices 
in the case of $ d =1 $:
\begin{equation*}
\begin{split}
P^T_t & = \frac{\lambda + t }{\lambda + 2T -t }
\frac{\theta^2 (\lambda + t)^2 + |Z^x_t|^2}
{\theta^2 (\lambda + 2T- t)^2 + |Z^x_t|^2}  \\
& \hspace{2cm} \times 
\frac{\theta^2 (1+c) (\lambda + 2T -t )^2 + c |Z^x_t|^2}
{\theta^2(1+c) (\lambda + t)^2 + c |Z^x_t|^2}. 
\end{split}
\end{equation*}
The model is a modification of the Cauchy TSMs given in \citep*{AkaTsu:07}.

\end{example}

Let $X_t $ be L\'evy process and $ T_t $ be a subordinator independent of $ X_t $. 
Recall that the process $ Z_t := X_{T_t} $ is usually called the subordination 
of $ X $ by the subordinator $ T $. Among the subordinations of the Wiener process, 
two classes are often used in the financial context; one is the {\em Variance Gamma} processes  and 
the other is {\em Normal Inverse Gaussian} processes, since they are analytically tractable (see \citep*{CoTa} 
and references therein). The subordinator of a Variance Gamma process
(resp. Normal Inverse Gaussian process) is a Gamma process (resp. Inverse Gaussian process). 
If we let the drift of the Wiener process be zero, then the densities of the subordinated processes 
satisfies the condition for Theorem \ref{mainTh}.

\begin{example}[Variance Gamma TSMs with Trace]
The Gamma subordinator $ T^G $ is a L\'evy process with 
\begin{equation*}
P (T^G_t \in dx )/dx = \frac{\gamma^{\eta t}}{\Gamma(\eta t)} 
x^{\eta t-1} e^{-\gamma x} 1_{ \{ x \geq 0 \} } , 
\end{equation*}
where $ \eta $ and $\gamma $ are positive constants. 
Then the heat kernel of the subordinated Wiener process 
$ W_{T_t^G} $ is
\begin{equation}\label{VGdensity}
u(t,x) := 
\int_0^\infty \frac{1}{\sqrt{2 \pi s}} e^{- \frac{x^2}{2s}}
\frac{\gamma^{\eta t}}{\Gamma( \eta t)} 
s^{ \eta t-1} e^{-\gamma s} \,ds. 
\end{equation}
From this expression we have
\begin{equation*}
u (t,0) = \sqrt{\frac{\gamma }{2 \pi }} 
\frac{\Gamma ( \eta t-\frac{1}{2})}{\Gamma (\eta t)}. 
\end{equation*}
The heat kernel (\ref{VGdensity}) can be expressed in terms of the modified Bessel functions
$ K_p $, $ p \in \mathbb{R} $, which has an integral representation
\begin{equation}\label{MBessel}
K_p (x) = \frac{1}{2} \left( \frac{x}{2} \right)^p 
\int_0^\infty e^{-t - \frac{x^2}{4t}} t^{-p-1} \,dt, \quad x > 0. 
\end{equation} 
Actually we have
\begin{equation*}
u (t,x) = \sqrt{\frac{2}{\pi}} 
\Gamma (\eta t)^{-1} 
\left( \frac{\gamma}{2}\right)^{-\frac{\eta t}{2}+\frac{1}{4}}
|x|^{ -\eta t + \frac{1}{2}} K_{ -\eta t + \frac{1}{2} }
\left( |x|\sqrt{2 \gamma} \right). 
\end{equation*}
\end{example}

\begin{example}[Normal Inverse Gaussian TSMs with Trace]
The heat kernel of the inverse Gaussian subordinator $ T^I_t $ is
\begin{equation*}
\rho_t (x) :=\frac{\eta t}{x^{3/2}} e^{2 \eta t \sqrt{\pi \gamma} } 
e^{-\gamma x - \pi \eta^2 t^2 /x }, \quad x > 0,
\end{equation*}
where $ \eta $ and $ \gamma $ are positive constants. 
The heat kernel of the subordinated 
Wiener process $ W_{T^I_t} $ is 
\begin{equation*}
u (t,x) :=  \int_0^\infty \frac{1}{\sqrt{2 \pi s}} e^{- \frac{x^2}{2s}}
\rho_t (s)\,ds. 
\end{equation*}
Applying (\ref{MBessel}), we obtain
\begin{equation*}
u(t,x) = \frac{1}{2 \pi} \frac{\eta t}{ y {\gamma}^3 } e^{ 2 \eta t \sqrt{\pi \gamma} } K_1 \left( \sqrt{ \gamma( 2x^2 + 4 \pi \eta^2 t^2)  } \right).
\end{equation*}
In particular, we have
\begin{equation*}
u(t,0) = \frac{1}{2 \pi} \frac{\eta t}{ y_0 {\gamma}^3 } e^{ 2 \eta t \sqrt{\pi \gamma} }
K_1 \left( 2 \eta t \sqrt{ \gamma \pi } \right).
\end{equation*}
\end{example}

\section{Appendix}\label{Ap:OU}

\begin{lemma}\label{A1}
For $ m, a, b \in \mathbb{R} $, $ v > 0 $,
and $ \mu < 0 $, it holds
\begin{equation}
\begin{split}
& \int_{a \leq x \leq b} e^{\mu x^2}\frac{1}{\sqrt{2 \pi v}} e^{-\frac{(x-m)^2}{2v}}dx \\
&=
\sqrt{\frac{\tilde{v}}{v}}
e^{\frac{\mu m^2}{1 - 2\mu v}}
\left(
\Phi(\frac{b-\tilde{m}}{\sqrt{\tilde{v}}})-\Phi(\frac{a-\tilde{m}}{\sqrt{\tilde{v}}}).
\right)
\end{split}
\end{equation}
where 
\begin{equation}\label{A2}
\tilde{m}:=
 \frac{m}{1 - 2\mu v}
 \ \ {\rm and} \ \  
\tilde{v} := \frac{v}{1- 2 \mu v}.
\end{equation}
\end{lemma}
\begin{proof}
\begin{equation}
\begin{split}
& \int_{a \leq x \leq b} e^{\mu x^2}\frac{1}{\sqrt{2 \pi v}} e^{-\frac{(x-m)^2}{2v}}dx \\
&
=
\sqrt{\frac{\tilde{v}}{v}}
e^{\frac{\mu m^2}{1 - 2\mu v}}
 \int_{a \leq x \leq b} \frac{1}{\sqrt{2 \pi \tilde{v}}} 
e^{-\frac{(x-\tilde{m})^2}{2 \tilde{v}}} dx \\ 
& 
=
\sqrt{\frac{\tilde{v}}{v}}
e^{\frac{\mu m^2}{1 - 2\mu v}}
\left(
\Phi(\frac{b-\tilde{m}}{\sqrt{\tilde{v}}})-\Phi(\frac{a-\tilde{m}}{\sqrt{\tilde{v}}})
\right). 
\end{split}
\end{equation}
\end{proof}

\subsection*{A proof of (\ref{E1}) with (\ref{E2}) and (\ref{E3})}

Notice that 
\begin{equation}
X_{T_\alpha -t} \big|_{X_t} \sim N(m_t,v_t)
\end{equation}
where 
\begin{equation}\label{A3}
m = X_t e^{-\mu (T_\alpha - t)} \ \ \text{and} \ \  
v = \frac{1}{2\mu} (1- e^{-2 \mu (T_\alpha - t)}).
\end{equation}

Put 
$$
a= - \sqrt{ \frac{1}{\mu} \log \frac{B}{A} } \ \ {\rm and} \ \  
b=   \sqrt{ \frac{1}{\mu} \log \frac{B}{A} }. 
$$
By Lemma \ref{A1}, we have 
\begin{equation}
\begin{split}
&\int_\mathbb{R} 
(A g(y) -B )^{+} P( X_{T_{\alpha}-t}^x \in dy \ | \ x=X_t ) \\
&=
A 
\sqrt{\frac{\tilde{v}}{v}}
e^{\frac{\mu m^2}{1 - 2\mu v}}
\left(
\Phi(\frac{b-\tilde{m}}{\sqrt{\tilde{v}}})-\Phi(\frac{a-\tilde{m}}{\sqrt{\tilde{v}}}).
\right)
- 
B
\left(
\Phi(\frac{b-m}{\sqrt{v}})-\Phi(\frac{a-m}{\sqrt{v}})
\right).
\end{split}
\end{equation} 
By (\ref{A2}) and (\ref{A3})
$$
\tilde{m}
= X_t e^{\mu (T_\alpha -t )}, \quad
\tilde{v}
= \frac{e^{2 \mu (T_\alpha - t)} -1}{2\mu},
$$
\begin{equation*}
e^{\frac{\mu m^2}{1 - 2\mu v}}
= e^{\mu |X_t^x|^2} = g (X_t^x), 
\end{equation*}
and 
\begin{equation*}
\sqrt{\tilde{v}/v}
= (1-2\mu v)^{-1/2} = e^{\mu (T_\alpha - t)}. 
\end{equation*}
With these expression we have (\ref{E1}). 
\qed 


\end{document}